\documentclass[english,preprint,onecolumn,aps,pre,eqsecnum,longbibliography,showpacs,showkeys,a4paper]{revtex4-1}
\usepackage[latin9]{inputenc}
\usepackage{color}
\definecolor{note_fontcolor}{rgb}{0.80078125, 0.80078125, 0.80078125}
\usepackage{babel}
\usepackage{amsmath}
\usepackage{amssymb}
\usepackage{esint}
\usepackage[unicode=true,
 bookmarks=true,bookmarksnumbered=true,bookmarksopen=true,bookmarksopenlevel=1,
 breaklinks=true,pdfborder={0 0 0},backref=false,colorlinks=true]
 {hyperref}
\hypersetup{
 pdfauthor={AINS}}

\makeatletter

\newenvironment{lyxgreyedout}
  {\textcolor{note_fontcolor}\bgroup\ignorespaces}
  {\ignorespacesafterend\egroup}

\@ifundefined{textcolor}{}
{%
 \definecolor{BLACK}{gray}{0}
 \definecolor{WHITE}{gray}{1}
 \definecolor{RED}{rgb}{1,0,0}
 \definecolor{GREEN}{rgb}{0,1,0}
 \definecolor{BLUE}{rgb}{0,0,1}
 \definecolor{CYAN}{cmyk}{1,0,0,0}
 \definecolor{MAGENTA}{cmyk}{0,1,0,0}
 \definecolor{YELLOW}{cmyk}{0,0,1,0}
 }
\numberwithin{equation}{section}
\numberwithin{figure}{section}
\numberwithin{table}{section}

\usepackage{graphicx}

\DeclareGraphicsExtensions{.png,.pdf,.eps}
\graphicspath{{pictures/}}

\makeatother

\begin{document}

\title{Classical Simulation of Double Slit Interference via Ballistic Diffusion}

\author{Johannes \surname{Mesa Pascasio}\textsuperscript{1,2}}

\email[E-mail: ]{ains@chello.at}

\homepage[Visit: ]{http://www.nonlinearstudies.at/}

\selectlanguage{english}%

\author{Siegfried \surname{Fussy}\textsuperscript{1}}

\email[E-mail: ]{ains@chello.at}

\homepage[Visit: ]{http://www.nonlinearstudies.at/}

\selectlanguage{english}%

\author{Herbert \surname{Schwabl}\textsuperscript{1}}

\email[E-mail: ]{ains@chello.at}

\homepage[Visit: ]{http://www.nonlinearstudies.at/}

\selectlanguage{english}%

\author{Gerhard \surname{Gr\"ossing}\textsuperscript{1}}

\email[E-mail: ]{ains@chello.at}

\homepage[Visit: ]{http://www.nonlinearstudies.at/}

\selectlanguage{english}%

\affiliation{\textsuperscript{1}Austrian Institute for Nonlinear Studies, Akademiehof\\
 Friedrichstr.~10, 1010 Vienna, Austria}

\affiliation{\textsuperscript{2}Institute for Atomic and Subatomic Physics, Vienna
University of Technology\\
Operng.~9, 1040 Vienna, Austria}

\affiliation{\vspace*{2cm}
}

\date{\today}
\begin{abstract}
Based on a proposed classical explanation, the quantum mechanical
``decay of the wave packet'' is shown to simply result from sub-quantum
diffusion with a specific diffusivity varying in time due to a particle's
changing thermal environment. The exact quantum mechanical intensity
distribution, as well as the corresponding trajectory distribution
and the velocity field of a Gaussian wave packet are therewith computed.
We utilize no quantum mechanics, but only familiar simulation techniques
for diffusion, e.g., finite differences or coupled map lattices (CML).%
\begin{lyxgreyedout}
\global\long\def\VEC#1{\mathbf{#1}}

\global\long\def\d{\,\mathrm{d}}

\global\long\def\e{{\rm e}}

\global\long\def\meant#1{\left<#1\right>}

\global\long\def\meanx#1{\overline{#1}}

\global\long\def\mpbracket{\ensuremath{\genfrac{}{}{0pt}{1}{-}{\scriptstyle (\kern-1pt +\kern-1pt )}}}

\global\long\def\pmbracket{\ensuremath{\genfrac{}{}{0pt}{1}{+}{\scriptstyle (\kern-1pt -\kern-1pt )}}}

\global\long\def\p{\partial}
\end{lyxgreyedout}

\end{abstract}


\maketitle

\section{Ballistic Diffusion}

We derive a solution for an anomalous diffusion equation with a time-dependent
diffusion coefficient $D_{{\rm t}}(t)=kt^{\alpha}$ of the general
form 
\begin{equation}
\frac{\partial P}{\partial t}=kt^{\alpha}\frac{\partial^{2}P}{\partial x^{2}}\;,\quad\alpha>0.\label{eq:tdde.1}
\end{equation}
Here, $t$ and $k$ denote the time and a constant factor, respectively.
$P(x,t)$ is the solution of Eq.~\eqref{eq:tdde.1}. We assume a
Gaussian distribution function as ansatz for $P$, 
\begin{align}
P(x,t)=\frac{1}{\sqrt{2\pi}\sigma}{\rm e}^{{\displaystyle -\frac{(x-x_{0})^{2}}{2\sigma^{2}}}}\;,\label{eq:tdde.2}
\end{align}
where $\sigma(t)$ denotes the standard deviation, i.e. the distance
from the mean value to the left or right interception point of the
distribution function $P$. For further considerations we put the
initial location $x_{0}$ in Eq.~\eqref{eq:tdde.2} into the coordinate's
origin, i.e., $x_{0}=0$. Then we calculate the partial derivatives
of $P$ using the short form $\dot{\sigma}$ for the time-derivative
of $\sigma(t)$,
\begin{align}
\frac{\partial P}{\partial t} & =-P\,\frac{\dot{\sigma}}{\sigma}+P\,\frac{x^{2}\dot{\sigma}}{\sigma^{3}}\;,\qquad\frac{\partial P}{\partial x}=-P\,\frac{x}{\sigma^{2}}\;,\qquad\frac{\partial^{2}P}{\partial x^{2}}=P\left(-\frac{x}{\sigma^{2}}\right)^{2}+P\left(-\frac{1}{\sigma^{2}}\right)\;,
\end{align}
then substitute these results into Eq.~\eqref{eq:tdde.1}, 
\begin{align}
\frac{P\dot{\sigma}}{\sigma}\left(\frac{x^{2}}{\sigma^{2}}-1\right) & =kt^{\alpha}\frac{P}{\sigma^{2}}\left(\frac{x^{2}}{\sigma^{2}}-1\right)\;,
\end{align}
and obtain 
\begin{align}
\frac{\sigma^{2}}{2} & =k\frac{t^{\alpha+1}}{\alpha+1}+\frac{c_{0}}{2}\;.\label{eq:tdde.5}
\end{align}
At this stage we introduce a result of \cite{Groessing.2010emergence}
which is an expression for the spreading of the Gaussian, 
\begin{equation}
\sigma^{2}=\sigma_{0}^{2}\left(1+\frac{D^{2}t^{2}}{\sigma_{0}^{4}}\right)\;,\label{eq:tdde.6}
\end{equation}
with $\sigma_{0}$ as initial standard deviation at time $t=0$. Note
that the diffusivity $D=\frac{\hbar}{2m}$ is constant for all times
$t$ and has to be distinguished from the diffusion coefficient $D_{{\rm t}}$.
Substitution of Eq.~\eqref{eq:tdde.6} into \eqref{eq:tdde.5} yields
$c_{0}=\sigma_{0}^{2}$ and 
\begin{align}
k\,\frac{2t^{\alpha+1}}{\alpha+1} & =\frac{D^{2}t^{2}}{\sigma_{0}^{2}}\;.\label{eq:tdde.7}
\end{align}
Eq.~\eqref{eq:tdde.7} can only be fulfilled by $\alpha=1$, so that
$k=\frac{D^{2}}{\sigma_{0}^{2}}\;.$ Hence, our time-dependent diffusion
coefficient becomes 
\begin{align}
D_{{\rm t}} & =\frac{D^{2}t}{\sigma_{0}^{2}}\;.\label{eq:tdde.8}
\end{align}
Finally, Eq.~\eqref{eq:tdde.1} becomes
\begin{align}
\frac{\partial P}{\partial t} & =D_{{\rm t}}\,\frac{\partial^{2}P}{\partial x^{2}}\label{eq:ballisticDE}
\end{align}
and turns out to be a ballistic diffusion equation because $\alpha=1$.

\section{Classical Simulation}

It is straightforward to simulate the diffusion process of Eq.~\eqref{eq:ballisticDE}
in a computer model. One approximates the diffusion equation by 
\begin{equation}
P[x,t+1]=P[x,t]+\frac{D[t+1]\Delta t}{\Delta x^{2}}\left\{ P[x+1,t]-2P[x,t]+P[x-1,t]\right\} \label{eq:1.34}
\end{equation}
 with space and time grid indices $x$ and $t$, respectively, and
an initial Gaussian $P(x,0)$ at $t=0$ with initial standard deviation
$\sigma_{0}$.

\section{Double Slit Interference}

The simulation of interference of two beams emerging from Gaussian
slits is established with the aid of a simulation as in Refs.~\cite{Groessing.2010emergence,Groessing.2011dice,Groessing.2012doubleslit,Grossing.2012quantum}.
To account for interference, we simply follow the classical rule for
the intensities $P_{{\rm tot}}:=R^{2}=\left|R_{1}\VEC k_{1}+R_{2}\VEC k_{2}\right|^{2}=R_{1}^{2}+R_{2}^{2}+2R_{1}R_{2}\cos\varphi=P_{1}+P_{2}+2\sqrt{P_{1}P_{2}}\cos\varphi$,
with $R\VEC k=R_{1}\VEC k_{1}+R_{2}\VEC k_{2}$ and $\varphi=m\Delta v_{x}\, x/\hbar$
. The trajectories are the flux lines obtained by choosing a set of
initial points at $y=0$. Two adjacent flux lines thereby define regions
of constant flux, i.e., $\int_{A}P\d A=\mathrm{const.}$, with $A$
being the cross section of a flux tube.

We have shown in \cite{Groessing.2012doubleslit,Grossing.2012quantum}
that it is possible to simulate a particle's trajectories behind a
double slit with usual classical simulation tools. The key is the
ballistic diffusion equation \eqref{eq:ballisticDE} which describes
the underlying physics by the time-dependent diffusion coefficient
\eqref{eq:tdde.8}.

%\bibliographystyle{utphys}
%\bibliography{../ains-reduced}
\providecommand{\href}[2]{#2}\begingroup\raggedright\endgroup

\end{document}